# On the possibility of breaking the heterodyne detection quantum noise limit with cross-correlation

E.A. Michael and F.E. Besser

Terahertz- and Astro-Photonics Laboratory, Department of Electrical Engineering, FCFM faculty, University of Chile, Av. Tupper 2007, Santiago, Chile

Corresponding author: E.A. Michael (emichael@ing.uchile.cl)

## Abstract

The cross-correlation sensitivity of two identical balanced photodiode heterodyne receivers is characterized. Both balanced photodiodes receive the same weak signal split up equally, a situation equivalent to an astronomical spatial interferometer. A common local oscillator (LO) is also split up equally and its phase difference between both receivers is stabilized. We show by semi-classical photon deletion theory that the post-detection laser shot noise contributions on both receivers must be completely uncorrelated in this case of passing three power splitters. We measured the auto- and cross-correlation outputs as a function of weak signal power (system noise temperature measurement), and obtain a cross-correlation system noise temperature up to 20 times lower than for the auto-correlation system noise temperature of each receiver separately. This is supported by Allan plot measurements showing cross-correlation standard deviations 30 times lower than in auto-correlation. Careful calibration of the source power shows that the auto-correlation (regular) noise temperature of the single balanced receivers is already very near to the quantum limit as expected, which suggests a cross-correlation system noise temperature below the quantum limit. If validated further, this experimentally clear finding will not only be relevant for astronomical instrumentation but also for other fields like telecommunications and medical imaging.

## I. Introduction

In astronomy the ultimate aim is to detect weakest signals over affordable integration times. Sometimes also highest angular and/or spectral resolution is necessary, adding the difficulties of interferometry and/or high-resolution spectroscopy. Due to the inherent quantum statistics [1] and the vacuum fluctuations of the electromagnetic field [2], fundamental limitations of sensitivity arise. Two radiation detection principles are competing here to achieve the higher signal-to-noise ratio (SNR) [3].

In direct (incoherent) detection the signal photons alone generate directly photoelectrons. For this, the detector is in integration mode and therefore slow. The sensitivity is then limited only by the counting noise of the signal photons detected from a thermal source in a measurement time interval $\Delta t$, $\delta n^2 = \bar{n}_S(\bar{n}_S + 1)$, but substantial post-detection amplifier noise adds to this. High spectral resolution is achievable only with bulky wavelength-dispersive optics in front of detector arrays [4], which is increasingly lossy towards higher resolutions.

In heterodyne (coherent) detection, the electromagnetic field to be detected, $P_s(\nu_S)$, is mixed on a fast detector (the mixer) with a strong monochromatic reference signal, the "local oscillator" (LO), $P_{LO}(\nu_{LO})$, down-converting the sidebands into the intermediate frequency (IF) band at $\nu_{IF} = |\nu_S - \nu_{LO}|$, preserving their phases. Therefore, the signal can be amplified in the very moment of detection so highly, by multiplication with the strong LO, that the impact of post-mixer IF-amplifier noise is eliminated. Unfortunately, this brings in fundamental quantum noise from the vacuum- or zero-point fluctuations (ZPFs) of the electromagnetic field, resulting in $h\nu/2$ of white noise power per Hertz and mode [2], [5]. Such can be formally regarded as emitted by a thermal source of "noise temperature" $T = h\nu/2k_B$.

Although ZPFs cannot be detected by a passive detector because they don't constitute real power [6], they still are supposed to necessarily add to the electromagnetic field at the input of a coherent receiver (and any phase preserving amplifier), and so contribute to the noise seen by it [7].

The mixing process contributes with another $h\nu/2$ for single sideband receivers and zero for double sideband receivers. Altogether, it results a smallest achievable system noise temperature (the quantum limit) of $T_{rec,SSB,min} = T_Q := h\nu/k_B$ for single-side band receivers and $T_{rec,DSB,min} = T_Q/2$ for double-side band receivers [7].

The mixing process in heterodyne detection can be understood as a parametric amplification from the two sidebands into the IF band. In phase-preserving amplification the requirement of an uncertainty relation $\Delta n \cdot \Delta \varphi \geq 1/2$ between the signal photon number and the absolute phase of the wave implies a minimum noise temperature contribution of the amplifier of $T_{rec,min} = T_Q/2$ [8], [9]. It can be shown already by using the simple relation between classical field strength and photon number in a mode, that the uncertainty relation between photon number and phase is closely related with quantum noise and electromagnetic zero-point fluctuations [10]. Note that $\Delta\varphi$ is absolute and does not mean a relative measure against the jittering LO phase. This implies for heterodyne interferometry that differences between two receivers, $\delta(\Delta\varphi) = \Delta\varphi_1 - \Delta\varphi_2$, are still allowed to be measured to higher precision than the uncertainty relation for a single receiver permits, in analogy to the formation of fringes in direct detection interferometry.

Very interestingly, $T_{rec,min} = T_Q$ results also when assuming within a semi-classical quantum theory the LO shot noise as the only noise source, ignoring the ZPFs. This coincidence appears to be not yet understood. It might be a result from ZPFs being present also in the oscillating mode of the laser cavity or in the circuitry of any oscillator based on amplifiers. This so-called standard quantum limit (SQL) noise arises then from the Poissonian probability distribution of photons in a single laser mode. It has a variance of $\delta n^2 = \overline{n}_{LO}$ around an average number $\overline{n}_{LO}$ during a measurement time interval $\Delta t$ [1]. The derivation of $T_{rec,min}$ with using the LO photon noise, as presented in many publications, e.g. in [11]-[16], is included here in the theory part. Such quasi-classical derivations are justified by the result Mandel et al. have established already in 1964, in that a full quantum field theoretical treatment is not necessary for the description of the photon detection process and a semi-classical description suffices [17].

According to Feldman et al. it depends on the conditions, which of both - the vacuum fluctuations or the LO photon noise - is the cause of the quantum limit for a single-mixer receiver [4], [18]. Single sideband configuration would make LO shot noise the source of the quantum limit, while double sideband configuration would make the zero point fluctuations its source. Note here that the receiver configuration used in the current investigation is double sideband.

As all the possible causes of the quantum limit should be equivalent and have indistinguishable effects, this paper uses the projection into the LO shot noise to develop a semi-classical quantum theory for a first explanation of the reported experimental result.

It seems to be not understood yet why direct detection should not see the noise of the vacuum fluctuations. On the other hand, why should then not exist a heterodyne receiver configuration avoiding to see these zero-point fluctuations? Whatever their nature is, it is too intriguing to find out how to bypass them in heterodyne detection. Applications for a receiver operating below the quantum limit would be ubiquitous: be it for laser interferometers in gravitational wave astronomy [19], or be it for imaging technologies in medicine (e.g. optical coherence tomography [20]) or in the life sciences (e.g. fluorescence microscopy [21]).

Especially in imaging astronomical interferometry highest possible visibility (cross-correlation) sensitivities are required to extend telescope baselines to the utmost [22]. Compared to the minimum object brightness necessary for single-telescope detection, in interferometry a higher brightness is necessary to detect increasingly weaker fringes at larger baselines for a sufficient sampling of the visibility curve, because the SNR suffers from the distribution of the available optical signal power to many detectors.

Since Albert Michelson demonstrated the first stellar interferometer a century ago, the technique of interferometry was steadily developed for high spatial resolution astronomy, using direct detection in the optical range (offering high spectral band width) [23], and heterodyne detection in the sub-millimeter range (offering high spectral resolution) [24]. For the purpose to maintain high spectral resolution, heterodyne interferometry was extended into the mid-infrared spectral range [25], [26].

The maximum possible sensitivity of heterodyne detection in cross-correlation between two receivers was tacitly assumed so far to be as well determined by the standard quantum limit. This might be another reason why for example in the ALMA interferometer array the sampling depth of the IF signal to be correlated was left at the 3 bits evaluated to be precise enough [27]. The here reported result from a detailed work on a new configuration of a correlation receiver puts this assumption now into question. However, the upgraded ALMA correlator will probably have 8 bits [28].

There is ample literature describing correlation receivers with the typical intention to suppress the uncorrelated thermal noise and amplifier gain fluctuations of the two parallel receivers, e.g. [29]. Surprisingly, seemingly none of them considers LOs with uncorrelated noise, all of them assume a single LO simply split up to feed both receiver chains.

In the present paper we exactly cover this case of uncorrelated LO noise on both receivers and demonstrate with it experimentally that the sensitivities of "traditional" balanced receivers and correlation receivers, both operating already near to the quantum limit, can be further improved substantially by combining both concepts into a so-called "balanced correlation receiver", which appears according to our experimental results to be capable of breaking the quantum limit. In fact, we measured about an order of magnitude increase of sensitivity (lower noise temperature) in cross-correlation compared to auto-correlation (single receiver). In the semi-classical picture (deletion of individual photons – no photon entanglement effects which are not necessary to beat the quantum limit [30]), described in the theory part, we can understand preliminarily that this not yet reported receiver configuration "correlates out" the statistically independent shot noise contributions of both LO signals against each other. But in case of a digital correlator, in order to do this, it is necessary to digitally resolve the LO shot noise sufficiently well, and so more ADC-bits are necessary than the ALMA correlator provides so far. In fact, we use 8-bit-ADCs.

The noise temperatures of the two balanced receivers (see Fig. 1) were measured carefully using a response plot over various source power levels (an extended hot-cold measurement), and were determined to be very near to the noise temperature quantum limit. This included the careful calibration of the spectral power density using an optical spectrum analyzer and several counter-checked NIST-traceable powermeters. The surprising but necessary conclusion is then that in cross-correlation the sensitivity reaches already clearly below the single-receiver quantum limit. To evaluate if this is not only a formal result but rather a real advantage of cross-correlation against auto-correlation, also Allan plots of all signal variances were made. The depth below the shot-noise limit we measure here (5-6 dB) is larger than was demonstrated previously with a photon number squeezed local oscillator in a single receiver (2-3 dB) [31]. The latter result gave already evidence that the SQL has also a contribution from the LO photon noise, not only from the influence of the vacuum state [5].

To provide a first explanation of this experimental finding, the work reported in this paper uses the essential assumption that the photoelectron fluctuations $\delta n_{el}$ (photocurrent) in a detector are strongly correlated with the radiation power fluctuations $\delta P_{rad}$, and therefore photon density fluctuations $\delta n$ seen by it in a time interval $\Delta t$. Moreover, this assumption is extended to fluctuations below the shot noise limit $\delta I_{el}^2 = 2eI_{el}\Delta f$, which is equivalent to $\overline{\delta P^2} = 2h\nu\bar{P}\Delta f$ on the radiation side, where $\Delta f = 1/2\Delta t$ is the fluctuation bandwidth. Otherwise it is not seen how the results of the current experiment and those of previous work [32] would be explainable. For this concept it is supportive that the density fluctuations of electrons in a dc-current flowing through a device tend to equal out due to the repelling force between the electrons and due to the fact that fermions cannot bunch into identical states, leading to a sub-Poissonian electron distribution [33], [34]. To be considered here is that a quantum of an electromagnetic mode cannot be localized, and therefore the single conversion event cannot be predicted for a perfectly monochromatic and hypothetically noiseless wave train of the LO with a photon number $\bar{n}_{LO}$. But because the probability of the absorption of a photon is proportional to the density of the photons, $n(t)$, the fluctuations of the photocurrent $\delta I(t)$ are correlated with $\delta n(t)$ down to the Nyquist time scale of $\Delta t_{Ny} = 1/2\Delta \nu$, where $\Delta \nu$ is the roll-off frequency of the detector. This does not lead to a conflict with quantum mechanics because the fluctuations $\delta n(t)$ are spectrally white.

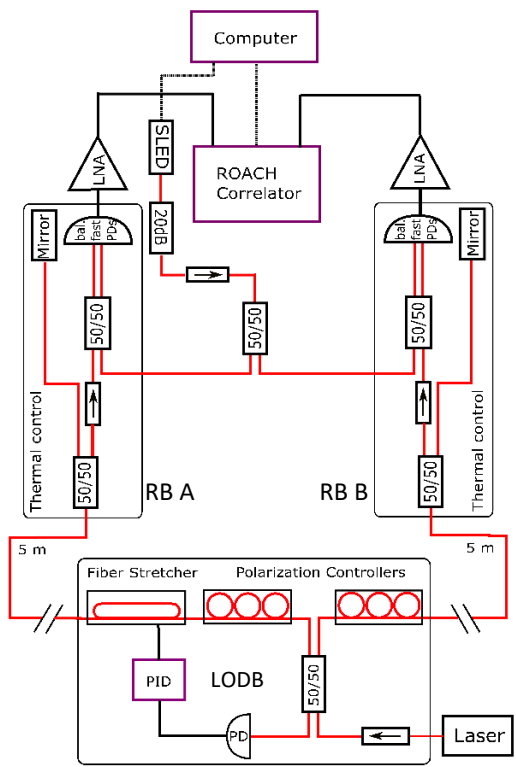

Fig. 1: Distribution of a laser LO signal to two independent receivers and subsequent measurement of the correlation. The Fano factor of the laser is $F \approx 10$ at 2 mW. As explained in the theory, after the balanced photodiodes it is unity, and both detected laser noise signals are completely uncorrelated. The laser phase difference at both balanced photoreceivers is locked to a precision of $\lambda/10$ over the Michelson interferometer fringes detected on the photodiode PD. As long as the phase variations are not too fast, they are captured by the correlator, so that the correlation amplitude is unaltered by slower phase drifts.

The paper is structured as follows: First we describe the experimental setup investigated with the intention to develop instrumentation for astronomical interferometry, and the effect we discovered with it. Then we show that this finding can be explained by a semi-classical quantum noise propagation theory (based on stochastic photon deletions), where we project the quantum limit into the shot noise of the local oscillator. Fundamental results from this formalism are derived in an appendix. This approach could be regarded as provisional. A full quantum-mechanical theory is desirable and may be developed later to completely understand the effect, for example using the concepts of [48] and [49]. This might then also grant additional insight into the origin of the quantum limit.

## II. Experimental setup

The experiment conducted is depicted in Fig. 1. It consists of two fiber-optic circuit units, the Local Oscillator (LO) distribution box (LODB, at the bottom center) and the receiver boxes (RB, at the center). A digital correlator was programmed onto a Reconfigurable Open Architecture Computer Hardware (ROACH1) platform. A fixed frequency fiber laser, a Koheras Adjustik (NKT Photonics), was used as a local oscillator (LO), working at 1556 nm with 1 kHz of linewidth and a thermal fine tuning capability of $\pm 0.5$ nm. Attenuated to a power or 3mW the laser has a Fano factor of about 10 [32].

The LODB contains an insulator at the LO input, to prevent standing waves due to back-reflections. A 50/50 fiber splitter distributes the laser power equally towards both receivers (in fact a tunable one to fine-adjust equal pump power to both balanced receivers to better than 5%), and redirects the fiber mirror reflections from there towards a slow photodiode (PD), on which interference fringes are formed (fiber-based Michelson interferometer). The PID control loop stabilizes the photodiode signal on the edge of one of these interference fringes through changing the fiber length in one of both arms with a fiber stretcher. It was tested to work up to a precision of $\lambda/10$ in our laboratory environment.

In the RB circuits a fiber splitter directs 50 % of the laser power towards the mentioned fiber mirror, and the other 50 % through an insulator, avoiding here any standing wave interaction between both balanced photodiode receivers assemblies. Those contain tunable fiber splitters in order to balance both photodiodes of each receiver to better than 5%. (We note here that splitting ratios of regular fiber splitters are in practice always a lot off the nominal 50/50 ratio needed for the balanced PDs, and in fact we could not select better fiber splitters than with a value of 40/60, i.e. $R_A$, $R_B = 0.4$ or $0.6$. This corresponds to a laser noise correlation coefficient of $c_{LO} \approx 0.07$ (see appendix B), and the first values we measured were in agreement with this. According to a derivation in the theory part such a deviation should raise the Fano factor after the balanced photodiode to F=1.4, considering a Fano-factor of the laser LO of F=10 at 2 mW [32].)

The balanced photodiode assemblies (Newport 1617-AC-FC) have a common-mode rejection of 25dB, a 3dB-roll-off frequency of 800 MHz, and include a trans-impedance amplifier of 11.5 dB gain (1A/W to 700V/W into 50Ω). After the balanced PDs the IF signals are amplified by another 40-45 dB [two LNAs of 35 dB gain and 1.4 dB NF, 0.02-3GHz BW, with a 20 dB attenuator in between both to prevent saturation in the second, and a 10 dB (later 5 dB) attenuator after them], before they are fed to the ADCs of the ROACH correlator.

The correlator is a ROACH1-board assembly (Reconfigurable Open Access Computing Hardware), containing a Xilinx FPGA, to which are attached two 8bit-iADCs of 3GS/s. The instrument was conceptualized and developed by the CASPER Group at Berkeley and fabricated by Digicom Electronics, Inc. The correlator model run on that was developed for a bandwidth of 800 MHz from the FX pocket correlator model available from the CASPER group. In order to reduce drifts from thermal instabilities, we have extended their model with a Dicke-switch for on/off-measurements as known from radio astronomy (see also [35]), which we run at 8 Hz using the reference from a chopper in case of using the halogen lamp, or at 1 Hz from the signal to switch on and off the SLED. Later we will use this scheme for position switching on sky between the target and a reference position, using a fiber actuator [35]. The signal power of a test source is split up equally and injected to each of the RBs. As test sources we used:

1.) A SLED (Exalos AG) with maximal single-mode fiber coupled power of 20mW. The driving current was set to 355mA, producing 13 mW, and was not touched any more since the band width depends on it in the range 52 - 60nm (6.4 - 7 THz). Two fiber attenuators separated by a fiber patch cord, an optical isolator, and a polarization controller were added which reduced the total power to 2.7 µW. The power integral $P_S$ over this band width we measured with a calibrated precision powermeter (Thorlabs PM100D). We used two InGaAs-photodiode power sensors (Thorlabs S154C/S155C) and checked also with a different powermeter model (Thorlabs PM20) which all gave similar power values within 10%. The calibration of all these powermeters is certified by Thorlabs as traceable to the NIST standard [36]. In fact, Thorlabs claims that the NIST-calibration at µW-levels is even more precise (direct against the power levels of the Mercury lamp after the monochromator filter band width) than at mW-levels (extrapolation).

At the 2.7 µW, the spectrum was characterized by an optical spectrum analyzer (Anritsu MS9740A). It is nearly of triangular shape with a maximum at 1540nm and a FWHM emission bandwidth of $\Delta v_S = 6.8$ THz. The spectral power density at 1556 nm is then near to 90% of the peak power, or 45 nW/nm. Additionally, the loss in the 50/50 power splitter towards the two receivers was determined to a few percent. But for the optical power calibration of the noise temperature measurement plot in Fig. 2c we leave these reductions aside to account for the 10% precision range of the used powermeters mentioned above. With the polarization controllers in the LODB the polarization overlaps of the SLED with the laser are maximized at each balanced photodiode. The (redundant) PC at the SLED is just used to check if the result behaves symmetric at both receivers.

As the large scaling factor of band width ratio between source and heterodyne band pass is precisely known, the input power per correlator channel can be accurately determined for the SLED. The measurement of the noise temperatures with the SLED can thus be regarded as very reliable and so those values are reported here. For the plot from which we extract them (Fig. 2c) we switched the source power between six very reproducible power values below the 2.7 µW level by passing it through a MEMS-based voltage-driven variable attenuator (Thorlabs V1550A).

2.) A halogen lamp (Ocean Optics Inc.) with radiation temperature 3500 K, from which estimated 5 mW total power were coupled into a 1mm-core multimode fiber (25.000 transversal modes at 1.55 µm). To this was connected a single-mode fiber receiving from it estimated 400 nW total power over the entire overlap pass band of fiber and optical powermeter (estimated to about 750nm). As this source had the least drifts, we recorded with it Allan plots to characterize the drift stability of the receiver near zero input power (about 0.5nW/nm, much smaller than the lowest measurement point with the SLED). Due to that low power, noise temperature measurements were not attempted with it.

3.) A fraction of 1% of the laser power split off, frequency shifted by around 55 MHz and attenuated into the fW-range. This feeds just a single correlator channel. Because the bandwidth of one channel is just 800 MHz /256 = 3.125 MHz, the power necessary for a noise temperature measurement is so small that it cannot be measured anymore directly. It requires the combination of separately calibrated attenuators and therefore suffers from a larger uncertainty in power calibration than the SLED. Therefore, noise temperatures obtained with this source were too uncertain and are not reported here.

### III. The receiver noise temperature

In order to enable the reader to follow the experimental results before reading the theory, it is better to introduce first how to measure the receiver noise temperature. The output power of a receiver as a function of optical source power (in Rayleigh-Jeans approximation proportional to the source temperature $T_S$) can be written as

$$P_{out}(T_S) = A(T_{rec} + T_S) = P_{out}(0) + \frac{\partial P_{out}}{\partial T_S} \cdot T_S \quad (1a)$$

where A is a constant. Thus, the receiver temperature can be also expressed as

$$T_{rec} = P_{out}(0) / \frac{\partial P_{out}}{\partial T_S} \quad (1b)$$

i.e. the receiver noise temperature is proportional to the output noise power at zero optical source input power (which is mainly laser noise, vertical axis section) divided by the slope of output noise power over signal input power. It is assumed that for the cross-correlation signal as a function of the source power we can perform the same determination of a receiver noise temperature.

In the radio and sub-millimeter wave range, normally a hot/cold load measurement is performed (Y-factor method) [37]:

$$T_{rec} = \frac{T_{S,hot} - Y T_{S,cold}}{Y - 1}, \text{ with } Y := \frac{P_{out,hot}}{P_{out,cold}} \quad (2)$$

where $P_{out,hot}$ and $P_{out,cold}$ are the noise output powers per frequency channel or the respective total power values. This is a special case where we have just two different load temperatures (emission powers) available (typical situation in submm-wave receivers where mostly $T_{S,hot} = 300$ K, and $T_{S,cold} = 77$ K), but in our case we can do a fit over many more source power values. In case of long wavelengths the equivalent radiation temperature of the load can be assumed to be proportional to the optical test source power (Rayleigh-Jeans approximation):

$$T_S = \frac{P_S}{k_B \Delta v_S} \quad (3a)$$

At shorter wavelengths it might be more precise to omit the Rayleigh-Jeans approximation and use the single-mode occupation number (see theory part)

$$\frac{P_S}{h v \Delta v_S} = n_S = \frac{1}{\exp(T_Q/T_S) - 1} > \frac{T_S}{T_Q} \quad (3b)$$

with $T_Q := hv/k$, giving higher values for the radiation temperatures. However, for our values, the difference for the source temperature is less than 10% for the considered values of optical input powers, so that we maintain the approximation in the following. Also, we observe a linear behavior over source power with this approximation.

At 1.55 μm wavelength, the ambient (300K) thermal emission background is practically zero, so that we can simplify eq. (2) with the cold load temperature being zero. Then, $P_{S,hot}$ can be inferred at any Y-factor along the linear measurement curve, or at the special point where $Y = 2$:

$$T_{rec} = \frac{T_{S,Y,hot}}{Y-1} = \frac{P_{S,Y}}{(Y-1)k_B \Delta\nu_S} = \frac{P_{S,Y=2}}{k_B \Delta\nu_S} \quad (4)$$

The critical part for the correct determination of the noise temperature is thus to determine accurately the spectral power density of the source at the LO frequency. In case of the SLED we could achieve it as described above: From the FWHM bandwidth, the total power and the position 10% off the maximum we determine it for the maximal of the plotted values of 2.7 μW to $P_{\nu,S} = P_S/\Delta\nu_S \cdot 0.9 = 45$ nW/nm. With a variable and calibrated MEMS-based fiber-attenuator we addressed then the various source power values in the plot of Fig. 2c with high reproducibility and precision.

## IV. Experimental results

In general, it is observed that the noise output power plotted over the signal input power follows a linear behavior for auto- and for cross-correlation. Furthermore, in cross-correlation the linear slope of the noise output power is the same as for auto-correlation. Second, it is observed that the noise residual (at zero input power) is up to 20 times weaker in cross-correlation than it is in auto-correlation. This both together gives a noise temperature up to 20 times less in cross-correlation.

The receiver noise temperature results using the SLED source are listed in Tab. 1 according to the plot in Fig. 2c. If we regard the DSB quantum limit, $T'_Q = T_Q/2 = h\nu/2k_B = 4.612$ K, then each receiver is at $4 - 5$ times $T'_Q$. Since in cross-correlation the input power for Y=2 is 12.8 - 21 times less, the receiver temperature would be there $0.28 - 0.45$ times $T'_Q$, clearly surpassing the quantum limit.

If we considered that in auto-correlation the injected power per receiver is half the value from the source due to the 50% power splitter, then the single-receiver noise temperature is in truth $2 - 2.5$ times $T'_Q$. The single-receiver noise temperatures should be elevated a bit above the quantum limit even for a balanced photodiode mainly because a photodiode quantum efficiency of $\eta = 0.75$ and sub-optimal LO pumping level [38], because the allowed maximal LO power is 1 mW per photodiode and thus the IF amplifier thermal noise still impacts the noise temperature (see theory part). However, the observed increase is larger than expected for a balanced photodiode. Note that the amplifier noise is purely thermal at room temperature (the quantum limit is very low in the IF band) and always must cancel out completely in cross-correlation, which is the only concept in conventional correlation receivers.

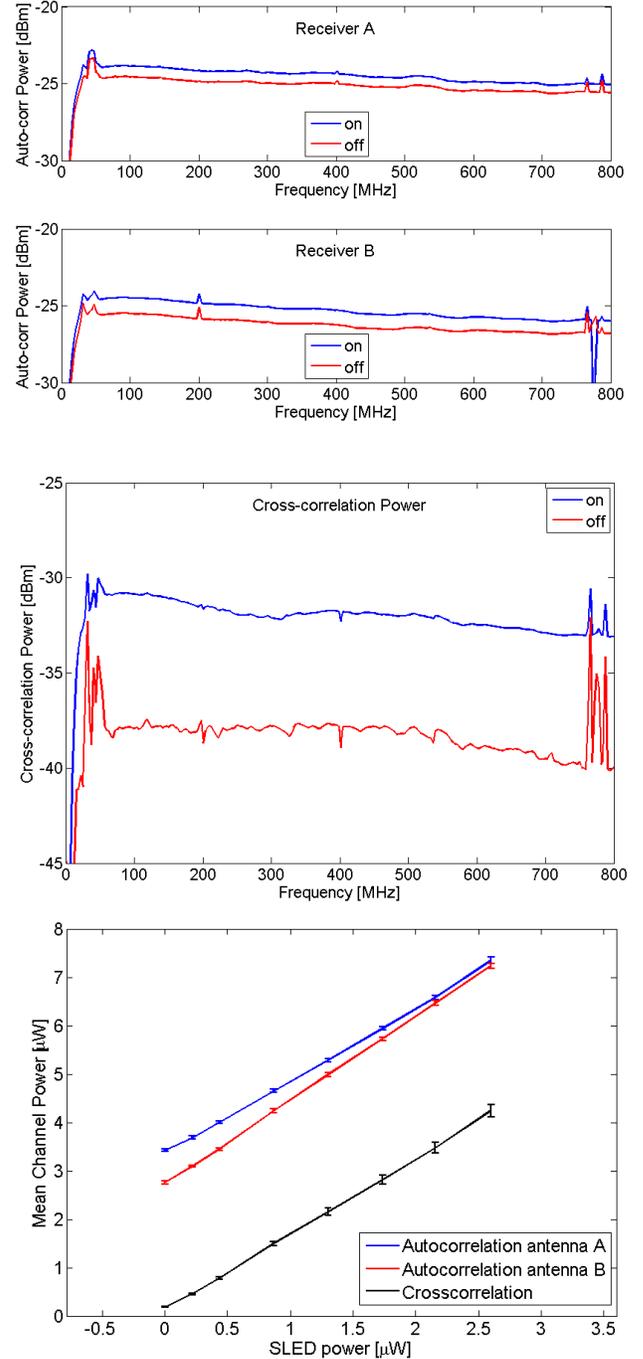

Fig. 2: a) Auto-correlation and b) cross-correlation mean ADC input power per channel after amplification of 66.5 dB, at 0.434μW total SLED power (integrated over 6.8 THz FWHM). The peaks near 770 MHz are cell phone interferences, those near 50 MHz are of unknown origin. c) Receiver noise temperature measurement: Same mean power per channel as function of total SLED power. The SLED power per spectral DSB channel (6.25MHz) is $9 \cdot 10^{-7}$ times the total power plotted, i.e. the maximal plotted optical input power is 2.3 pW per channel.

| SLED | $P_{Y=2}$ (plot) | $P_{v,Y=2}$ 1) | $T_{rec,DSB}$ | $\times T_Q'$ |
|---|---|---|---|---|
| auto-corr. A | 2.3 µW | 2160 fW | 25,040 K | 5.43 |
| auto-corr. B | 1.6 µW | 1530 fW | 17,740 K | 3.85 |
| cross-corr. AB | 0.15 µW | 135 fW | 1,565 K | 0.34 |
| 1) per $\Delta v_{DSB} = 6.25$ MHz, $T_Q' = T_Q/2 = h\nu/2k_B = 4{,}612$ K | | | | |

*Tab. 1: Summary of noise-temperature results with the SLED. The input power for each Y=2-point is given for the DSB channel band width (6.25 MHz, 256 channels within 800 MHz SSB).*

The SLED results may still contain unnecessary common-mode drift noise because of the chopping method we chose. The electronics board of the SLED contains a control input pin to switch it on and off and we used this to chop it. However, the transition times are rather slow, so that the highest chopping frequency is 1 Hz. Thus, the chopping duty cycle may have drifted slightly.

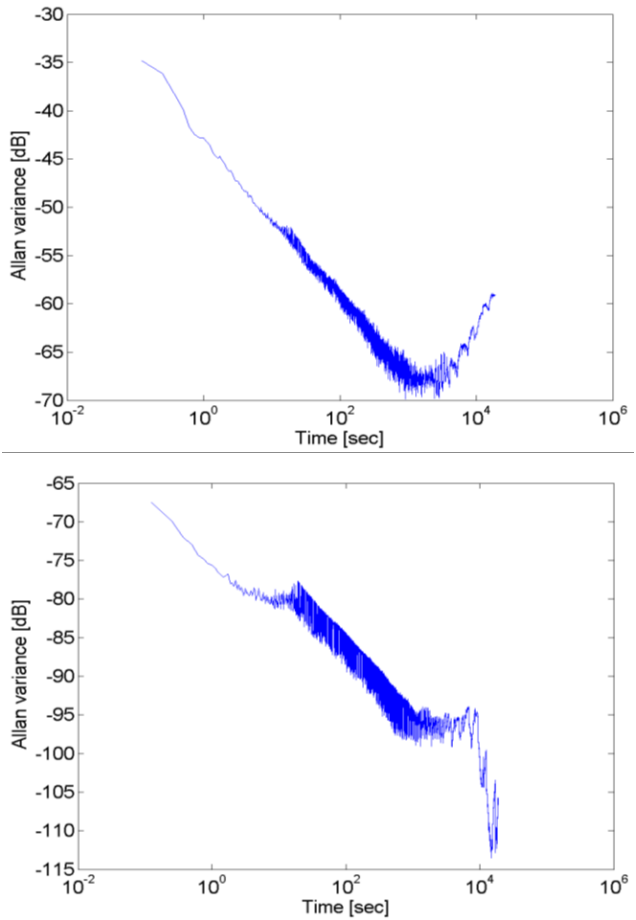

*Fig. 3: Allan-plots of the variance as a function of the integration time, of a) the auto-correlation of receiver A (B very similar), b) the cross correlation between both receivers at 400 nW total halogen lamp power coupled into the single-mode fiber, resulting in about 7 pW in the 1.6 GHz DSB bandwidth (about 16 fW per DSB channel – near to zero power in plot 2c). The time series of the data we recorded for these plots was about 10 hours.*

The main doubt about these cross-correlation noise temperature measurements is about whether the residual cross-correlation noise level itself would not be important but rather the standard deviation from the mean noise level would determine the sensitivity. If the absolute value of the standard variation in cross-correlation would be the same as in auto-correlation [39]. Then the reduced noise level would not help. However, the experimental prove finally arises from the absence of this expected strength of cross-correlation fluctuations, which can be seen from the Allan plots in Fig. 3, measured with using the fiber-coupled halogen lamp. It was optically chopped with a mechanical chopper at 8 Hz. The Allan plot code was developed from [40], and its performance was cross-checked against simple codes readily available.  The result visible from the plots in Fig. 3 is that the Allan variance $\sigma^2$ of the cross-correlation is 30 dB below that of the auto-correlation, so that the rms-error is 31.6 times smaller. Because the output noise power at zero input power levels is 10-20 times weaker in cross-correlation the relative fluctuations there are still 1.5-3 times smaller. One can state then that the whole single-receiver response curve is scaled down in a self-similar way to smaller optical input power levels.

The cancellation of the laser shot noise of both receivers in cross-correlation is expected to depend on the precision of calculation in the correlator. It should be maximized at optimal sampling of the laser shot noise which would occur at a certain amplification (see appendix C). However, then signals larger than the laser shot noise should run already into saturation.

The experimental correlation coefficient using a 10 dB attenuator after the amplifier chain (that used to record the plots in Fig. 2, total amplification 56.5 dB) resulted in $c_{LO} \approx 0.06$. For this observed noise cancellation value we estimate to have used only less than 10% of the dynamic range of the ROACH-ADCs, whereas the optimum is at 50% of this range (see appendix C). For this optimum value the shot noise is expected naively to cancel to 99%, leading to a correlation coefficient of $c_{LO} = 0.01$.

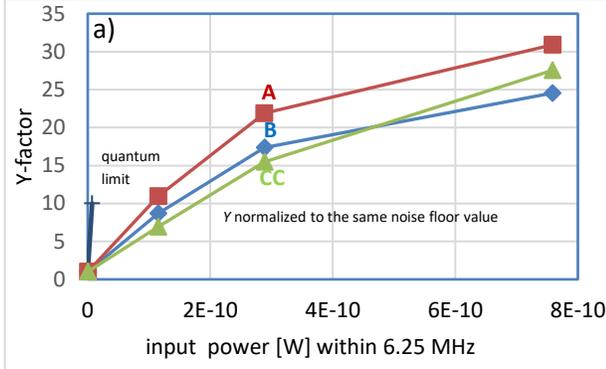

| Last attenuator value [dB] | AC ZIN [dBm] | CC ZIN [dBm] | Δ ZIN [dB] | $c_{LO} = \left.\dfrac{P_{CC}}{P_{AC}}\right|_{P_s=0}$ |
|---|---|---|---|---|
| 10 | -25.08 | -37.28 | -12.20 | 0.060 |
| 5 | -19.95 | -33.22 | -13.27 | 0.047 |
| 0 | -15.52 | -26.62 | -11.09 | 0.078 |

Tab. 2: Dependence of the laser noise suppression (or expected noise temperature improvement in cross-correlation) in dependence of the calibrated ADC input power levels of auto- (both receivers averaged) and cross-correlation (ZIN = noise power at zero optical input signal, only laser).

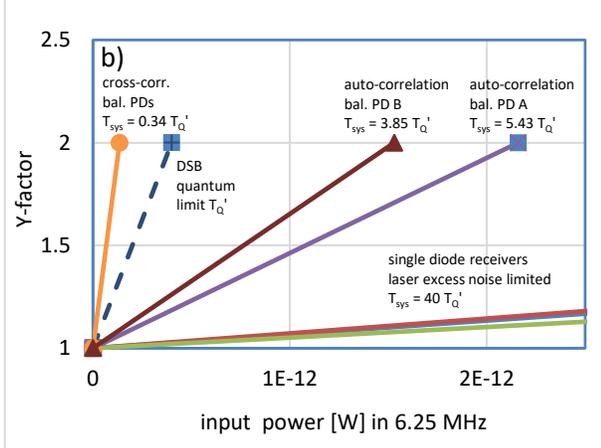

Fig. 5: a) Single photodiode receivers Y-factor plot as a function of the signal strength up to 0.8 nW per channel (with saturation at the highest input power). b) Y-factors of both receiver-types in comparison, normalized to the respective different zero-input-power noise-power levels.

As listed in Tab. 2, we ran the plot of Fig. 2c also with amplifications 5 and 10 dB higher (reducing the last attenuator to 5 dB and then to 0 dB). The laser correlation coefficient $c_{LO}$ (the ratio of cross-correlation to auto-correlation at zero source power) should then drop, and along with it also the expected cross-correlation noise temperature according to eq. (1b). This resulted experimentally to be counteracted by saturation, so that the optimum is "only" at $c_{LO} = 0.047$ (21 times better noise temperature). This means that if we would have receiver A equally good as receiver B, we would reach $T_{CC} = 0.2 \times T_Q'$.

To show that with single-mixer receivers cancellation of the laser noise cannot be achieved, a control measurement was made with a setup in which the local oscillator is split equally just one time towards two single photodiodes (Thorlabs DET01CFC, quantum efficiency $\eta = 0.75$). The output of them is subsequently correlated. The second input of the fiber splitter accepts the signal, so that the total assembly resembles a balanced photodiode. But the outputs of both photodiodes are not just simply subtracted from each other but rather correlated. (Mere unplugging of one of the photodiodes in each of the balanced photodiode assemblies is not working because then an extreme DC-bias will be imposed to the input of the pre-amplifiers in them.) With this, we again measured the auto- and cross-correlation of the amplifier outputs as a function of the optical input, and clearly the residual laser noise power was the same in cross-correlation as in auto-correlation (Fig. 5a). In Fig. 5a this noise temperature measurement is plotted. In auto-correlation the noise temperatures are $40 \times$ away from the DSB quantum limit, and the cross-correlation has the same zero-input noise level and therefore the same unreduced noise-temperature like both receivers have separately. The Allan plot of the cross-correlation is then observed to have similar parameters as those for auto-correlation, i.e. the same start- and minimum $\sigma^2$ and the same Allan time. This measurement also verifies that the correlator is working correctly. In Fig. 5b we compared the low source power end of the single PDs with the balanced PDs measurement. All plotted Y-factors (IF output noise ratio $Y := P_{IF,on}/P_{IF,off}$, $Y = 1$ means input off, just background noise) are plotted relative to the respective (different) zero-input noise floors.

It has to be commented here that the single-PD result seems to contradict the experiment of Hanbury Brown and Twiss (HBT), who state that for coherent light no (cross-)correlations between directly detected photon density (intensity) fluctuations should occur (they had a single atomic emission line extracted through a monochromator) [41]. However, in heterodyne detection we cross-correlate fluctuations of the E-fields (first-order coherence function) rather than the intensities (second order coherence function). According to the beam splitter photon bunching effect, also called Hong-Ou-Mandel effect (HOM) [42], it could be understood that a scarce signal photon may then tend to pair up with one of the many laser carrier photons which is nearby enough in time so that both leave the splitter through the same port. This way the same source photon statistics as in direct detection may be observable also in heterodyne detection. It additionally may be that the result we observe is only valid for the excess noise component of the laser signals, and if we would use a laser at the shot noise limit, then no correlation may be observable also in case of single photodiodes with a single splitter, similar to the result of Hanbury Brown and Twiss for direct detection which was theoretically explained by Glauber [43]. However, no information was given about the Fano factor $F$ of the radiation sources used in the HBT experiments. Probably that was considerably above unity due to the thermal nature of the line emitters. In future, our measurement results should be verified with the current laser replaced by one operating very close to the shot noise limit. However, this would require creating a laser power of 120 mW without amplification. It may also be realizable by filtering the current laser through an extremely narrow-band ASE filter or reducing the excess noise by a special circuit [32].

## V. Theory
### A. General considerations

In order to understand provisionally the reported findings, a semi-classical theory was developed. We derive here the pre-detection signal-to-noise ratios (SNRs) in auto- and in cross-correlation, which would be the upper bounds of any SNRs possible. The post-detection SNRs take into account additional deteriorating post-detection noise sources, but these are secondary for the comparison of auto- and cross-correlation.

In the following, $\Delta v_S$ is the spectral interval in which the detector or receiver is sensitive, i.e. for a heterodyne receiver it is also the IF bandwidth. $\Delta f = 1/2\tau$ is the fluctuation bandwidth (in post-detection the integration filter bandwidth), where $\tau$ is the light integration (exposure) time over which we average the radiation noise, or in post-detection the time over which the readouts of a spectral channel are averaged. The respective achievable integration times depend on the Allan times of the system.

### B. Thermal source and laser LO radiation

The total number of modes a detector interacts with is given from the cavity mode theory by the mode number per volume $\Delta V$, solid angle $\Delta\Omega$ and spectral interval $\Delta v$, $\Delta m_{cav,ang} = 2(v^2/c^3)$ (for two polarizations) as:

$$\Delta M_{cav,ang} = 2(v^2/c^3)\Delta V\Delta\Omega\Delta v \qquad (5)$$

$$= 2(1/\lambda)^2\Delta A\Delta\Omega\Delta v(\Delta L/c) =: 2\Delta M_A \Delta M_L$$

where $\Delta M_A = \Delta A\Delta\Omega/\lambda^2 \geq 1$ is the number of transversal modes the detector "sees" (étendue, antenna theorem), and $\Delta M_L := \Delta v\Delta t$ is the number of longitudinal traveling wave modes arriving at it (from one direction) in $\Delta t = \Delta L/c$. Then the detected source power in one polarization is the average energy in a mode, $hv\,\bar{n}_S$, multiplied with the total number of modes arriving, and divided by the time $\Delta t$ in which their waves pass the antenna. For heterodyne detection (a single transversal mode and a single polarization) the detected power is:

$$P_S = hv\,\bar{n}_S\,\Delta M_L/\Delta t = hv\,\bar{n}_S\,\Delta v_S =: P_{S,v}\Delta v_S \qquad (6)$$

in which the so-called "mode occupation number" $\bar{n}_S$ is for thermal continuum light fields (Planck radiation law)

$$\bar{n}_S = 1/(e^{hv/kT} - 1) \qquad (7)$$

In an astronomical situation a source can be point-like, i.e. much smaller than the receiver beam. Then an overlap factor $\eta_S \ll 1$ has to be written in place of $\Delta M_A$, which is, however, not important for the discussion in this paper, and therefore is omitted in the following.

For the laser LO we have:

$$P_{LO} = hv\,\bar{n}_{LO}\,\Delta v_{LO} \qquad (8)$$

with $\bar{n}_{LO} \gg 1$, and the laser line width being very small $\Delta v_{LO} \ll \Delta v_S$, depending on the laser coherence length $L_{coh} = c/\Delta v_{LO}$.

### C. Radiation noise

To obtain the signal field rms power fluctuations received by a detector, the energy fluctuations $(\delta E_m)^2 = (hv)^2(\delta n_m)^2$ are equal in each mode interacting with the antenna and are statistically independent:

$$(\delta E_{total})^2 = \sum_{m=1}^{\Delta M_{cav,ang}}(\delta E_m)^2 \qquad (9)$$

$$= \Delta M_{cav,ang}(\delta E_m)^2 = 2\,\Delta M_A\,\Delta M_L(hv)^2(\delta n)^2$$

For a single polarization state and a single (fundamental) mode it is

$$(\delta E_{total})^2 = \Delta v\Delta t\,(hv)^2(\delta n)^2 \qquad (10)$$

The rms power fluctuations become

$$\delta P = \sqrt{(\delta E_{total})^2}/\Delta t \qquad (11)$$

With the fluctuation bandwidth $\Delta f := 1/2\Delta t$ ($\Delta t =$ integration time) we have for the variations:

$$\delta P = hv\sqrt{(\delta n)^2\,\Delta v/\Delta t} = hv\sqrt{2\,(\delta n)^2\Delta v\Delta f} \qquad (12)$$

In case of thermal continuum radiation, eq. (7), the mode occupation number fluctuations can be calculated to [1]

$$(\delta n_S)^2 = \bar{n}_S(\bar{n}_S + 1) \qquad (13a)$$

The laser power fluctuations in the shot-noise limit are white and therefore follow Poisson statistics. This means that the photon number $\bar{n}_{LO}$ seen in a certain time interval $\Delta t$ has a variance given by $(\delta n_{LO})^2 = \bar{n}_{LO}$ [1]. For smaller excess noise levels, predominantly arising from ASE, we assume that this excess noise is still nearly white and the variance can still be approximated by

$$(\delta n_{LO})^2 = F\bar{n}_{LO} \qquad (13b)$$

with $F \geq 1$, the Fano-factor [32]. With the integration time, $\Delta t = 1/2\Delta f$ (where $\Delta f$ is the fluctuation bandwidth), an equivalent expression is $\overline{\delta P^2} = 2h\nu\bar{P}\Delta f$. The relative intensity noise of a laser is defined as $RIN \coloneqq \overline{\delta P^2}/\bar{P}^2 = \int S(f)df$, in which the shot noise spectral power density is $S_{SN}(f) = 2h\nu/\bar{P}$, which is white from first principles.

### D. Pre-detection SNR

It makes sense to define the pre-detection (optical) signal-to-noise ratio (SNR) as

$$SNR_{pre} = 1/RIN = \bar{P}_{opt}^2 / \overline{\delta P_{opt}^2} \qquad (14)$$

because due to $I_{det} = \Re P_{opt}$ (with $\Re = \eta e/h\nu$ the responsivity in photoconductive detectors) and $P_{el} = ZI_{det}^2$, this is consistent with the post-detection (electronic) signal-to-noise ratio:

$$SNR_{post} = \bar{P}_{el}/\sqrt{\overline{\delta P_{el}^2}} = \bar{I}_{el}^2 / \overline{\delta I_{el}^2} \qquad (15)$$

Eq. (14) does not make a big difference to the conventional way to define it, since with $f \coloneqq g^2$ and $\Delta f = 2g\Delta g$ we have $SNR_{sq} = f/\Delta f = g^2/2g\Delta g = g/2\Delta g = SNR/2$. Considering now heterodyne, we can write the single-receiver beat signal of the total resulting E-field before detection (expectation value of the total photon number is $\propto P \propto |E|^2$):

$$P_{tot}(t) = P_S + P_{LO} + 2\sqrt{P_S P_{LO}} \cdot \cos(\omega_{IF}t + \varphi_S) \qquad (16)$$

so that the rms average of the variations which enter into the heterodyne signal is

$$P_{het} = \sqrt{2P_S P_{LO}} \qquad (17)$$

Note that $P_{het}(t)$ oscillates positive and negative with an amplitude proportional to the signal E-field phasor projected to the LO E-field phasor:

$$\widehat{\boldsymbol{P}}_{het}(t) \propto E_S e^{i\varphi_S} E_{LO} e^{i\omega_{IF}t} =: \hat{E}_S(\omega_{IF},t)\hat{E}_{LO} \qquad (18)$$

The mode occupation number fluctuations within the IF-band and the laser sidebands are now to be considered for the thermal signal and the quasi-monochromatic (coherent) LO power fluctuations. The heterodyne rms power fluctuations could be calculated analog to eq. (10), but because we need for the cross-correlation their phasors, we write first:

$$\delta\widehat{\boldsymbol{P}}_S(t) = h\nu\sqrt{\bar{n}_S(\bar{n}_S + 1) \cdot \Delta\nu_S\, 2\Delta f} \cdot \hat{\boldsymbol{n}}_S(t) \qquad (19a)$$
$$\delta\widehat{\boldsymbol{P}}_{LO}(t) = h\nu\sqrt{F\bar{n}_{LO} \cdot \Delta\nu_{LO}\, 2\Delta f} \cdot \hat{\boldsymbol{n}}_{LO}(t) \qquad (19b)$$

$\hat{\boldsymbol{n}}_S(t)$ and $\hat{\boldsymbol{n}}_{LO}(t)$ are fast-changing complex Gaussian random phasors related to the signal and laser noise, respectively. They describe equal probabilities for all phase angles and Gaussian probability distribution for the amplitude around the mean (2-D bell-shape), and are uncorrelated. Therefore

$$|\delta P_{tot}|^2 = |\delta P_S|^2 + |\delta P_{LO}|^2 \qquad (19c)$$
$$= (h\nu)^2[\bar{n}_S(\bar{n}_S + 1)\Delta\nu_S + F\bar{n}_{LO}\Delta\nu_{LO}]2\Delta f$$

$F = (\delta n_{LO})^2/\bar{n}_{LO} > 1$ describes the white amplified spontaneous emission (ASE) excess noise of the laser. $F = 1$ in the IF can be achieved by a balanced coherent mixer, which is introduced shortly in the following. (The principle and semi-classical quantum theory of a balanced photodiode (BPD) mixer is explained in detail in appendix B.) With balanced photodiodes, $\delta P_S$ and $\delta P_{LO}$ are subtracted out at IF frequencies because they are common-mode on both photodiodes. A statistical particle deletion theory [44] was used to model photon noise propagation, which is explained in appendix B. It is in accordance with simple quantum optical experiments with photon counting detectors and beam splitters [45]. The statistical behavior of (detected) photons at the beam splitter (in front of the detectors), following Poisson statistics, is not subtracted out but adds up because they are differential-mode on the photodiodes. Therefore, beam splitter noise at the shot noise level (partition noise) arises [46], with a random phasor $\hat{\boldsymbol{n}}_{LO,BS}(t)$ which is statistically independent from the original laser excess noise random phasor ($\hat{\boldsymbol{n}}_{LO,BS}(t)$ is uncorrelated to $\hat{\boldsymbol{n}}_{LO}(t)$). Such beam splitter noise is in agreement with a full quantum mechanical description of the two outputs with injection of a coherent input into just one of the ports and vacuum fluctuations into the other [47]. It has to be cleared finally also how the Hanbury Brown and Twiss statements are compatible with it [41], [42], [43]. It follows:

$$\delta\widehat{\boldsymbol{P}}_{S,BS}(t) = h\nu\sqrt{\bar{n}_S\,\Delta\nu_S\,2\Delta f} \cdot \hat{\boldsymbol{n}}_{S,BS}(t)$$
$$= \sqrt{h\nu\, P_S\, 2\Delta f} \cdot \hat{\boldsymbol{n}}_{S,BS}(t) \qquad (20a)$$
$$\delta\widehat{\boldsymbol{P}}_{LO,BS}(t) = \sqrt{h\nu\, P_{LO}\, 2\Delta f} \cdot \hat{\boldsymbol{n}}_{LO,BS}(t) \qquad (20b)$$

We have now instead of eq. (19c) with $F = 1$:

$$|\delta P_{tot}|^2 = |\delta P_{S,BS}|^2 + |\delta P_{LO,BS}|^2 \qquad (20c)$$
$$= (h\nu)^2[\bar{n}_S\Delta\nu_S + \bar{n}_{LO}\Delta\nu_{LO}]2\Delta f$$

and the pre-detection SNR of heterodyne detection is in a balanced detector scheme:

$$SNR_{het} := \left(\frac{P_{het}}{\delta P_{tot}}\right)^2 = \frac{2P_S P_{LO}}{|\delta P_{S,BS}|^2 + |\delta P_{LO,BS}|^2}$$

$$= \frac{2(h\nu\,\bar{n}_S\,\Delta\nu_S)(h\nu\,\bar{n}_{LO}\,\Delta\nu_{LO})}{(h\nu)^2\,[\bar{n}_S\Delta\nu_S + \bar{n}_{LO}\Delta\nu_{LO}]2\Delta f}$$

$$= \bar{n}_S\,\frac{1}{1 + \frac{P_S}{P_{LO}}}\,\frac{\Delta\nu_S}{\Delta f} \to \bar{n}_S\,\frac{\Delta\nu_S}{\Delta f} \quad (21)$$

the last for $P_{LO} \gg P_S$. Furthermore, the partition noise process is statistically independent at each of the balanced receiver's power splitters, i.e. $\hat{n}_{LO,BS,1}(t)$ is uncorrelated to $\hat{n}_{LO,BS,2}(t)$, which is important for the following.

### E. Pre-detection SNR for the cross-correlation of two receivers

At each of the receivers, the instantaneous heterodyne power phasor for a specific IF is, including noise phasors:

$$\hat{P}_{het}(t) = \sqrt{2P_S P_{LO}}\,\hat{s}(t) + \delta P_S\,\hat{n}_S(t) + \delta P_{LO}\,\hat{n}_{LO}(t) \quad (22)$$

with $\hat{s}$ the normalized IF signal phasor. Transformed into a frame of the complex plane rotating at the beat (IF) frequency $\omega_{IF}$, $\hat{s}_0$ is constant in direction and length, whereas $\hat{s}(t) = \hat{s}_0 e^{i\varphi_j(t)}$, with $\varphi_j$ the signal ($j = 1,2$) E-field phase relative to the LO E-field phase.

According to the Wiener-Khinchin theorem (see [35], p. 56-58, and derivation in appendix D) the cross-correlation power spectrum between two receivers in real time is equal to the product of the Fourier-transforms of both single-receiver heterodyne power time-signals. This in turn, for real-time calculations, is approximated by calculating the fast-Fourier-transforms (FFT) of both single-receiver heterodyne power signals with a shortest possible integration time of $2T$ centered around $t$ ($\pm T$, large enough for the lowest IF, in our case 20 MHz), multiplying the FFTs of both receivers and averaging that "fast" product over a longer time (FX-correlator). Because of

$$P_{CC,T}(\omega_{IF}, t) \propto \tilde{E}_{1,T}(\omega_{IF}, t) \cdot \tilde{E}_{2,T}^{*}(\omega_{IF}, t) \quad (23)$$

and eq. (18) (heterodyne power phasor proportional to the signal E-field phasor), we can write:

$$\overline{\tilde{P}_{het,1,T}(\omega_{IF}, t) \cdot \tilde{P}_{het,2,T}^{*}(\omega_{IF}, t)} = 2\,P_{CC,T}(\omega_{IF}) \cdot P_{LO} \quad (24)$$

where $\overline{(\ )\cdot(\ )}$ means the time-average of the product of the spectral components. Note that this is analog to the single-receiver (auto-correlation) expression (17) for the heterodyne power. Assuming equal strengths of signal and noise terms, respectively, at both receivers we have:

$$\overline{\tilde{P}_{het,1,T}(\omega, t) \cdot \tilde{P}_{het,2,T}^{*}(\omega, t)} =$$
$$P_S P_{LO}\,\overline{\hat{s}_1 \hat{s}_2^{*}} + |\delta P_S|^2\,\overline{\hat{n}_{S1}\hat{n}_{S2}^{*}} + |\delta P_{LO}|^2\,\overline{\hat{n}_{LO1}\hat{n}_{LO2}^{*}} \quad (25)$$
$$=: S_{CC} + N_{CC}$$

where the time-averaged correlation terms between the random phasors are:

$\overline{\hat{s}_1 \hat{s}_2^{*}} =: \gamma$   Normalized correlation of the signals (visibility), the quantity to be measured by an interferometer (the spatial coherence function of the signal over the projected baseline of the telescopes). In case of a single signal split for two receivers, i.e. for a zero-baseline spatial interferometer, $\gamma = (\boldsymbol{p}_1 \cdot \boldsymbol{p}_2)e^{i\Delta\varphi}$, depending on the angle between the polarizations and the phase difference.   (26a)

$\overline{\hat{n}_{S1}\hat{n}_{S2}^{*}} =: \gamma \cdot c_S$   Correlation of the signal noise at both receivers. In the test setup it should be $c_S = c_{LO}$ from symmetry reasons at the power splitters.   (26b)

$\overline{\hat{n}_{LO1}\hat{n}_{LO2}^{*}} =: c_{LO}$   correlation of the laser noise at both receivers. This is to be discussed in the following.   (26c)

Other cross correlation terms, like $\overline{\hat{n}_{S1}\hat{n}_{LO2}^{*}}$ or $\overline{\hat{s}_1\hat{n}_{S2}^{*}}$, we assume to go to zero fast enough as the integration time increases. It is desirable to derive how the error of these terms depend explicitly on the integration time T, but this was not included so far.

The SNR for cross correlation is therefore in the analog picture eq. (20c) and (21):

$$SNR_{CC} := \left(\frac{S_{CC}}{N_{CC}}\right)^2 = \frac{2\gamma P_S P_{LO}}{\gamma|\delta P_S|^2 + c_{LO}|\delta P_{LO}|^2}$$

$$= 2\gamma(h\nu\bar{n}_S\Delta\nu_S)(h\nu\bar{n}_{LO}\Delta\nu_{LO})\,/$$
$$(h\nu)^2[\gamma\bar{n}_S(\bar{n}_S + 1)\Delta\nu_S + c_{LO}F\bar{n}_{LO}\Delta\nu_{LO}]\,2\Delta f$$

$$= \gamma\bar{n}_S\,\frac{1}{c_{LO}F + \frac{[\gamma\,\bar{n}_S(\bar{n}_S + 1)]\,\Delta\nu_S}{\bar{n}_{LO}\,\Delta\nu_{LO}}}\,\frac{\Delta\nu_S}{\Delta f}$$

For $\delta P_{LO} \gg \delta P_S$ and $P_{LO} \gg P_S$ this gives, compared to (21):

$$SNR_{CC} = \frac{\gamma}{c_{LO}}\,\frac{\eta_S \bar{n}_S}{F}\,\frac{\Delta\nu_S}{\Delta f} = \frac{\gamma}{c_{LO}} \cdot SNR_{het} \quad (27)$$

Therefore, the SNR in cross correlation with completely uncorrelated LO signals but correlated signal ($\gamma = 1$) is expected to be much better than for single receivers. If the laser LO signals are completely correlated, then it is $SNR_{CC} = SNR_{AC}$, as also verified by the reported control experiment with single photodiodes as receivers. Then the partition noise in the two LO signals from the single power splitter is opposite identical, $\delta n_2 = -\delta n_1$.

The RIN of two independent lasers is obviously completely uncorrelated, but for interferometry their phases have to be locked, and so their RINs may result to have mutually dependent parts due to the phase lock circuits.

In our interferometer we avoided such an unclear situation by making the approach to use just one laser and distribute it over fiber power splitters to the receivers. Then the LO signals at the two receivers would be totally correlated (as in the mentioned control experiment), if we would not remove that correlation by using balanced photodiode receivers. Such cancel out any noise of the LO and, from symmetry reasons, also of the signal, because this noise is in common mode after the balancing power splitter. If we assume that the photons are detected after this power splitter either in the one or in the other photodiode, a power splitter noise must result which is anti-correlated (differential mode) on the two photodiodes and therefore adds up, recreating exactly the shot noise level (see appendix B). But because the two balancing power splitters at the two receivers are completely independent, also the new shot noise is they generate. To explain this fact, and also other noise experiments [32], it is necessary to assume that the photon fluctuations are translated, within the time resolution of the fastest detected IF frequency and with an appreciable correlation, into photoelectron fluctuations.

**F. Post-detection SNR**

Several deteriorating factors appear in post-detection, like the quantum efficiency $\eta < 1$ of the detector and amplifier noise. From the pre-detection (optical) SNR we get then using the post-detection (electrical) signal and noise powers $S_{post}$ and $N_{post}$, respectively, the post-detection signal-to-noise ratio, according to (11) and (12):

$$SNR_{post} = \frac{S_{post}}{N_{post}} = \frac{Z\Re^2(S_{pre})^2}{Z\Re^2(N_{pre})^2 + N_{el}}$$

$$= \frac{SNR_{pre}}{1 + \frac{N_{el}}{Z(\Re N_{pre})^2}} =: \frac{SNR_{pre}}{1 + NR} \quad (28)$$

where $\Re \coloneqq \eta\, e/h\nu$ is the responsivity (in A/W), $Z$ is the amplifier chain input (load) impedance.

Here, we use the assumption that the radiation fluctuations are detected, i.e. translated with high correlation into post-detection electronic noise, even at the sub-shot noise level. This works according to the Burgess variance theorem [44], see appendix B:

$$(\delta n')^2 = \eta(1-\eta)\bar{n} + \eta^2 \cdot (\delta n)^2 \quad (29a)$$

with $\bar{n}' = \eta \cdot \bar{n}$, and $\bar{n}$ being the number of photons in the mode, and $\bar{n}'$ being the number of photoelectrons generated. It is then with eq. (6) and (12):

$$(\delta I)^2 = (e/h\nu)^2 (\delta P')^2$$
$$= (e/h\nu)^2 \cdot 2(h\nu)^2 \Delta\nu \Delta f (\delta n')^2$$
$$= \Re^2 \cdot [2h\nu P \Delta f(1-\eta)/\eta + (\delta P)^2] \quad (29b)$$

Therefore, the electronic noise power spectral density of the post-detection amplifier chain, projected to the detector, has the following form:

$$N_{el} = \left(k_B T_{sys,ampl} + 2e(I_{det,dark} + (1-\eta)I_{ph})\right)\Delta f \quad (29c)$$

The deteriorating noise ratio factor results as:

$$NR_{het} =$$
$$\frac{\left(k_B T_{sys,ampl} + Z \cdot 2e(I_{det,dark} + (1-\eta)I_{ph})\right)\Delta f}{Z\Re^2(h\nu)^2[(\bar{n}_S(\bar{n}_S+1))\Delta\nu_S + F\bar{n}_{LO}\Delta\nu_{LO}]2\Delta f}$$

$$\approx \frac{(k_B T + Z \cdot 2eI_{dark})}{2Z\Re^2 h\nu \, FP_{LO}} + \frac{1-\eta}{F\eta} \quad (30)$$

With sufficiently high LO power and high quantum efficiency we can achieve $NR_{het} < 1$, so that heterodyne detection can be regarded as almost ideal, i.e. $SNR_{post} \rightarrow SNR_{pre}$. For the real case of limited LO-power the SNR is deteriorated.

**G. Auto- and cross-correlation in post-detection**

In our setup, the instantaneous IF voltage signals are digitized after an amplification of about 67 dB and the spectrum is calculated by fast Fourier transform (FFT) from data chunks $1/\Delta\nu_{ch}$ long (here 0.3 µs), therefore calculating the quasi-instantaneous signal power spectral density, here with resolution $\Delta\nu_{ch} = \Delta\nu_S/256$. The noise-contaminated (time-varying) spectral voltage phasor before amplification is:

$$\widehat{V}_{het}(t) = \sqrt{Z\, 2Z\Re^2 P_S P_{LO}}\, \widehat{s} \quad (31)$$
$$+ \sqrt{Z \cdot ZF2e\Delta\nu_S \cdot \Re P_{LO}}\, \widehat{n}_{sh}(t)$$
$$+ \sqrt{Z \cdot 4kT\Delta\nu_S}\, \widehat{n}_{th}(t)$$
$$= Z\sqrt{2\Re P_{LO}\Delta\nu_S} \times \left[\sqrt{\Re\widetilde{P}_S}\, \widehat{s} + \sqrt{Fe}\, \widehat{n}_{sh}(t) \right.$$
$$\left. + \sqrt{2kT/(Z\Re P_{LO})}\, \widehat{n}_{th}(t)\right]$$

$\widehat{s}$ is the signal phasor, which should be rather stable in direction and length, while $\widehat{n}_{sh}(t)$ and $\widehat{n}_{th}(t)$ are again fast-changing Gaussian random variables related to the laser noise and the post-detection thermal electronic noise, respectively, but here related to voltages instead of powers.

The heterodyne post-detection IF rms signal power results as

$$S_{het,el}(\omega) = \frac{1}{2Z}\overline{|\widehat{V}_{het}(\omega,t)|^2} = 2Z\Re^2 P_S P_{LO} \quad (32)$$

The noise power at the frequency $\omega_{IF}$ is determined by the shot noise of the laser and the thermal noise of the detector coupling impedance to the IF-amplifier chain.

$$N_{het,el} = Z\delta I^2 = Z\,F\,2e\Delta\nu_S\,\Re P_{LO} + kT\Delta\nu_S \quad (33)$$

Then the SNR is

$$\begin{aligned}SNR_{het,el}(\omega_{IF}) &= \frac{S_{het,el}}{N_{het,el}} \\ &= \frac{2Z\Re^2\,P_S(\omega_{LO}\pm\omega_{IF})\,P_{LO}}{Z\,F\,2e\Delta\nu_S\,\Re P_{LO} + kT\Delta\nu_S} \\ &= \frac{\Re\widetilde{P}_S(\omega_{LO}\pm\omega_{IF})}{F\cdot e + kT/(2Z\Re P_{LO})}\end{aligned} \quad (34)$$

since $P_S = \widetilde{P}_S \Delta\nu_S$ (continuum source). At sufficiently high laser power the noise is dominated completely by the shot noise of the laser:

$$SNR_{het} \to \frac{\Re\widetilde{P}_S}{F\cdot e} = \frac{\widetilde{P}_S}{NESPD}$$

where

$$NESPD := \frac{Fh\nu}{\eta} \quad (35)$$

is the laser-noise limited noise equivalent spectral power density (NESPD) of heterodyne detection. Note that its dimension is W/Hz, and not W/Hz$^{\frac{1}{2}}$ as in the NEP. If we define a system noise temperature over $NESPD := kT_{sys}$, then

$$T_{sys,AC} = \frac{F}{\eta}\frac{h\nu}{k} = \frac{F}{\eta}T_Q \quad (36)$$

and $T_Q$ is the quantum limit of the receiver noise temperature. Then we can write $SNR_{het} = T_S/T_{sys,AC}$, and it must be equal to $Y$ in eq. (2).

In the more realistic case of intermediately high laser power it is:

$$T_{sys} = \frac{F}{\eta}T_Q + \frac{T}{2Z\Re^2 P_{LO}} \quad (37)$$

In our test case, with the laser power at $P_{LO} = 1$ mW at each of the two balanced photodiodes ($F = 1$) (regarding a single receiver), and with $\eta = 0.75$, at $T = 300$ K this gives $T_{sys} \approx 1.7 \cdot T_Q$.

The noise temperature itself should not be confused with the variations it has. Those are given by the radiometer formula which is valid for the long-wavelength and/or high-temperature range ($\bar{n}_S \gg 1$): According to

$$\delta P_S(t) = h\nu\sqrt{\bar{n}_S(\bar{n}_S+1)\cdot\Delta\nu_S\,2\Delta f}$$

$$\approx h\nu\bar{n}_S\Delta\nu_S\sqrt{\frac{2\Delta f}{\Delta\nu_S}} = P_S/\sqrt{\Delta\nu_S\Delta t}$$

and using eq. (4) for the hot temperature, $\delta T_{rec} \propto \delta P_{S,hot}$ and $T_{rec} \propto P_{S,hot}$ it is

$$\delta T_{rec} = T_{rec}/\sqrt{\Delta\nu_S\Delta t} \quad (38)$$

In a heterodyne interferometer, two signals like eq. (31) are correlated by multiplying the two instantaneous FFT power signals (FX-correlator, see Wiener-Khinchin theorem, appendix D). The cross-correlation power between two voltage signals of that kind is

$$\begin{aligned}CC_T(\omega_{IF}) &= \frac{1}{2Z}\overline{V_{het,1}(\omega_{IF},t)V_{het,2}^*(\omega_{IF},t)} \\ &= 2Z\Re P_{LO}\Delta\nu_S \\ &\quad\cdot\Big[\Re\widetilde{P}_S\,\overline{\widehat{s}_1\widehat{s}_2} + eF\,\overline{\widehat{n}_{sh1}(\omega_{IF},t)\widehat{n}_{sh2}(\omega_{IF},t)} \\ &\quad + \frac{2kT}{(Z\Re P_{LO})}\overline{\widehat{n}_{th1}(\omega_{IF},t)\widehat{n}_{th2}(\omega_{IF},t)}\Big]\end{aligned} \quad (39)$$

where the first term is the signal and the second and third term are noise terms. Mixed terms like $\overline{\widehat{n}_{sh1}(t)\widehat{n}_{th2}(t)}$ are assumed to be zero anyway. With the definitions of (26) for the correlation of the different phasors we get

$$CC_T(t) = 2Z\Re P_{LO}\Delta\nu_S \cdot [\Re\widetilde{P}_S\cdot\gamma + eF\cdot c_{LO}(\omega_{IF},t)] \quad (40)$$

where LO-correlation coefficient depends on time and gives a noisy part if the integration (averaging) time is finite. The signal-to-noise ratio is therefore

$$SNR_{CC} = \frac{\Re\widetilde{P}_S\cdot\gamma}{eF\cdot c_{LO}} =: \frac{\widetilde{P}_S}{NESPD_{CC}} = \frac{T_S}{T_{sys,CC}} \quad (41)$$

is linked over the noise-equivalent cross-correlation spectral power density $NESPD_{CC} =: k_B T_{sys,CC}$ to the cross-correlation system noise temperature

$$T_{sys,CC} = \frac{c_{LO}}{\gamma}\frac{F}{\eta}T_Q = \frac{c_{LO}}{\gamma}T_{sys,AC} \quad (42)$$

Therefore, cross-correlation has a sensitivity advantage over auto-correlation if the LOs of both receivers are not noise-correlated, a situation we have created experimentally.

## VI. Conclusions and discussion

It was verified experimentally that in cross-correlation of two heterodyne balanced receivers (four detectors) weaker signals can be measured than using a single balanced receiver (two detectors). A statistical particle deletion picture [44] is used to explain this result provisionally. First it is used to explain how balanced receivers reach quantum limited sensitivity. Any laser excess noise inclusive underlying shot noise is cancelled by the output subtraction after each balanced photodiode. Then, both balanced photodiode's power splitters create new laser shot noise through partition noise which is not cancelled. Such is locally and spontaneously created at each power splitter and therefore uncorrelated between both balanced receivers. This fact is exploited to create a cross-correlation of the laser noise from both balanced receivers a factor of up to 20 smaller than obtained in auto-correlation for each receiver alone, and this already with amplifying the signal much less than the optimum calculated (see appendix C). Therefore, with the slope of IF output versus optical signal input maintained from auto-correlation, a noise temperature a factor of up to 20 less results.

In comparison, the cross-correlation of two single photodiode receivers after one power splitter showed no increased sensitivity compared to the auto-correlation of each of them. Because the LO signal is split up just once here, the two split LO signals carry opposite identical fluctuations, and so both are perfectly correlated, so that there is no difference in sensitivity between auto- and cross-correlation.

However, this situation has not been the case for most heterodyne interferometers so far operated. The LOs in systems like ALMA, VLA or ISI are independent but phase-locked to a central reference and so uncorrelated in amplitude noise. Therefore, the effect should be detectable there as well. However, in ALMA the digital resolution might be too low for resolving the effect, but in ISI, analog cross-correlation was used. Therefore, while the quantum theory of heterodyne detection was validated thoroughly during the development of sub-mm heterodyne receivers in the past 40 years, it is not clear why the here reported effect was not discovered earlier in interferometric systems.

Regarding single-telescope receivers, nowadays the development of balanced sub-mm receivers reaches maturity, so that concepts combining balanced detectors with sideband separation are being conceived. Therefore, the reported effect is likely to be scrutinized soon by others, also at sub-mm wavelengths, and possibly applied soon.

## VII. Outlook

The present experimental result encourages to investigate, regardless of frequency, dual-balanced-mixer correlation receivers for single telescopes in continuation of the development of single-balanced-mixers. This should be also very interesting for submm- and terahertz-astronomy, since semi-classical theory predicts a sensitivity increase as well for this wavelength range. The result also suggests to extend heterodyne technology for interferometry into the mid-infrared, for example for the currently discussed Planet Formation Imager (PFI) [22], since the cross-over to direct detection appears to be shifted by the new extra-sensitivity factor towards higher frequencies or even to vanish completely in favor of heterodyne detection, if comparing both detection methods at equal bandwidths. The result could also be interesting for the discussion of upgrades for existing facilities [28]. In order to continue consolidating the result also theoretically, a full quantum mechanical approach should be developed for the correlation of two balanced mixers, considering for example the approaches of Collett et al. [48] and Zmuidzinas [49].

## VIII. Acknowledgements

This work was realized in the framework of the CONICYT Chile grants ALMA 31080020, 31090018, and 31110014, 31140025, and Quimal 150010. Support was also given through the Chilean Center for Excellence in Astrophysics and Associated Technologies (Project PFB 06), and through Fondecyt 1151213. Furthermore, we acknowledge helpful comments from J. Stutzki, University of Cologne, Germany.

## IX. Appendices

**Appendix A**
**Balanced photodiodes in heterodyne mode**

The current signals at the two photodiodes of a balanced mixer are [50]:

$$I_1(t) = \frac{\eta_1 e}{h\nu}\{R \cdot P_S + T \cdot P_{LO} + 2\sqrt{RTP_SP_{LO}} \cdot \cos[(\omega_1 - \omega_2)t + \varphi_1 - \varphi_2 - \pi/2]\}$$

(A1)

$$I_2(t) = \frac{\eta_2 e}{h\nu}\{T \cdot P_S + R \cdot P_{LO} + 2\sqrt{RTP_SP_{LO}} \cdot \cos[(\omega_1 - \omega_2)t + \varphi_1 - \varphi_2 + \pi/2]\}$$

The laser and signal dc-power levels cannot be detected by a BPD in direct detection because they are common mode (equal) on both PDs and so subtract out for $R = T$. However, the IF signals have a phase difference of $\pi$ between both diodes because of the conservation of signal, LO and total powers ($|r_{LO}|^2 + |t_{LO}|^2 = 1$, $|r_S|^2 + |t_S|^2 = 1$, and $P_1 + P_2 = P_{LO} + P_S$). Therefore, both IF signals add up after subtraction and are fully received. For a rigorous treatment it should also be discussed elsewhere whether $\delta \boldsymbol{P}_S$ and $\delta \boldsymbol{P}_{LO}$ would both be received through the mixing process, whereas $\delta \boldsymbol{P}_{S,BS}$ and $\delta \boldsymbol{P}_{LO,BS}$ would cancel out in heterodyne detection (inverse behavior to direct detection).

## Appendix B
### Laser noise propagation and de-correlation

In our setup we use a single laser of which the excess noise signals must be common-mode after the first distribution power splitter. Therefore, balanced photodiodes with the related 50/50-power splitters in front are necessary in order to replace the laser noise in each of the two receivers by the power-splitter partition noise [44], [46], which is undistinguishable from shot noise and completely uncorrelated to the original laser noise. Therefore, the laser noise at both receivers would be uncorrelated.

The crucial assumption of this paper is that we can achieve in principle that $c_{LO} \to 0$. For this we exploit that the normalized Gaussian random noise phasor $\hat{n}_1$ (and $\hat{n}_2 = -\hat{n}_1$) generated at a power splitter A in a balanced photodiode assembly A is statistically independent from that generated at a second power splitter B in another balanced photodiode assembly B. To derive this, we consider the mentioned semi-classical particle deletion model in the following: When a light signal is propagated it is either attenuated, split, or amplified. The question is what happens in these cases with the imprinted noise starting with the signal. In our case of the distribution of the LO laser signal, all these signal parts are finally detected by the different detectors at the telescopes (the heterodyne mixers). In that case an individual LO photon is finally detected only at one of these mixers, and so has to behave at the intermediate power splitters like a particle which is either transmitted or reflected, but not both at the same time which would be the behavior of a wave-like signal. This means that with the detection (absorption) at a particular detector we destroy all the other possible states of the overall probability amplitude wave function.

Interference at the detectors is produced only if two or more parts of the split signal are recombined again later, letting their (probability-, E-field-) amplitudes add up (interfere) like waves (Mach-Zehnder or Michelson interferometer). Then "fringes" are produced at all the detectors as a function of the path length difference. Because then the path of an individual photon cannot be determined anymore, it takes effectively both pathways at the same time and interferes with itself at the detector like a wave (see [45] for a collection of instructive experiments on these concepts).

The electron statistics in an electric current, e.g. that of a photodetector, or the photon statistics in a quasi-monochromatic laser beam, are both Poissonian in the limit of minimum possible noise, the shot noise limit, i.e. the probability to meet an electron or a photon is $p(n) = \bar{n}^n e^{-\bar{n}}/n!$ in a certain time interval, of which the variance is $(\delta n)^2 = \bar{n}$. If we assume a minimal time interval still resolvable by the detector, we have $\Delta t = 1/B$, where $B$ is the bandwidth of the detector (and which normally is regarded as the same as $\Delta \nu_S$), and so we have $\delta i^2 = 2e\bar{I}B$ (current shot noise). The temptation is now to believe that the photons before the detector are absorbed exactly at the same times the photoelectrons are ejected in the photodetector. However, in a monochromatic laser beam, according to the uncertainty relation $\Delta E \cdot \Delta t \geq \hbar/2$, photons cannot be localized at all, even if it is a travelling-wave mode. Rather, it can be only stated that the probability $p_e(t)$ to absorb (and so detect) a photon, and so to generate a photoelectron, is proportional to the cycle-averaged intensity just in front of the detector,

$$p_e(t) \propto |\overline{E} + \delta E(t)|^2 \propto n_{ph}(t) \quad (B1)$$

up to the frequency of the bandwidth of the detector. Support for this was worked out by the author with an optical/electrical noise interferometer experiment [32]. (Unfortunately, in that paper from space restrictions it was not formulated clearly enough in the point that not all the photons but just the photon fluctuations must obviously be detected 1:1 in order to explain the experiment.) Support for this is also given by [33] and [34].

This is equivalent to the fact that, additionally to the monochromatic spectrum of the laser, a broadband background spectrum must be superimposed, the relative intensity noise (RIN) background, which reduces in the limit without excess noise to the white shot noise background. Because that photon noise background is white, there is no conflict with the uncertainty relation to assume that these photons are localized and are converted 1:1 to the correspondent photoelectron fluctuations. The surprising result of the reported investigations is obviously that this holds down into the sub-shot noise regime.

The average number of photons, $\bar{n}$, should oscillate with the frequency $\omega$ of the field according to the probability interpretation, and therefore can be regarded as a phasor rotating in the complex plane. We can transform into a rotating system in which the signal phase is constant. Additive to the signal phasor is the noise phasor $\delta n(t)$, so that in total $n(t) = \bar{n} + \delta n(t)$ (see Fig. 1 b).

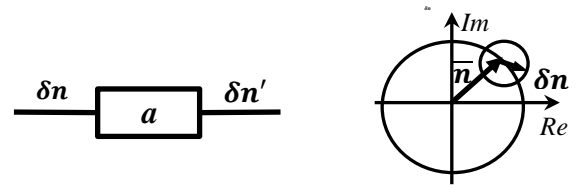

*Fig. B1: a) A passive attenuator. b) Photon number phasor $\bar{n}$ and noise phasor $\delta n$.*

According to Perina et al [44], attenuation is the stochastic deletion of photons, and so is not just the multiplication of the input fluctuations by an absorption factor $a < 1$, but also the addition of a new stochastic phasor. This results in the following expression for the propagation of the stochastic fluctuation phasor $\boldsymbol{\delta n}$:

$$\boldsymbol{\delta n'} = \sqrt{a(1-a)\bar{n}} \cdot \boldsymbol{\hat{n}_a} + a \cdot \boldsymbol{\delta n} \quad (B2)$$
$$\bar{n}' = a \cdot \bar{n}$$

with $\boldsymbol{\delta n'} = \delta n' \cdot \boldsymbol{\hat{n}'}$ and $\boldsymbol{\delta n} = \delta n \cdot \boldsymbol{\hat{n}}$. $\boldsymbol{\hat{n}}$, $\boldsymbol{\hat{n}_a}$ and $\boldsymbol{\hat{n}'}$ are normalized stochastic and statistically independent (orthogonal) noise phasors so that $\overline{\boldsymbol{\hat{n}_i} \cdot \boldsymbol{\hat{n}_j}} = \delta_{ij}$, where the bar means integration over time. Their amplitudes have a Gaussian probability distribution centered at zero, and their phases have a flat probability distribution (bell shape over the complex plane). Consider that the previous expression can be used also in the evaluation of correlation between two noise signals, as $\int \boldsymbol{\hat{n}_i}(t) \cdot \boldsymbol{\hat{n}_j}(t+\tau)dt = \delta(\tau)\delta_{ij}$. Fig. B1 shows a power splitter which can be regarded, with photon-counting detectors behind it, either in transmission or reflection also as a device which deletes photons, writing $a' = R$ and $a'' = T$, but with the additional condition $R + T = 1$. This conservation of photons requires that in the case that we have two detectors after the power splitter we can detect the photon either in transmission or in reflection, but not in both at the same time [45]. Therefore, the fluctuations generated at the splitter in transmission and reflection are related by $\widetilde{\delta n'} = -\widetilde{\delta n''}$, or for the stochastic phasors with $\boldsymbol{\hat{n}'} = -\boldsymbol{\hat{n}''}$, and therefore the two noise signals after the power splitter in front of the balanced photodiode are:

$$\boldsymbol{\delta n'} = \sqrt{TR\,\bar{n}} \cdot \boldsymbol{\hat{n}} + T \cdot \boldsymbol{\delta n} \quad (B3)$$
$$\boldsymbol{\delta n''} = -\sqrt{TR\,\bar{n}} \cdot \boldsymbol{\hat{n}} + R \cdot \boldsymbol{\delta n}$$
$$\bar{n}' = T \cdot \bar{n}$$
$$\bar{n}'' = R \cdot \bar{n}$$

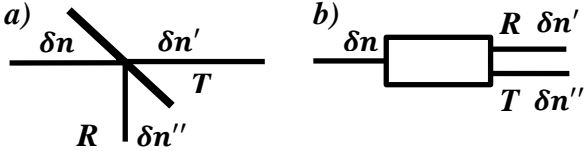

Fig. C2: a) A free-space, and b) a fiber power splitter

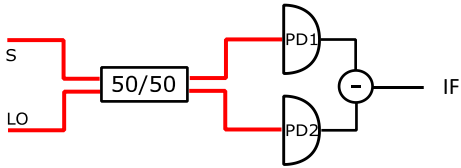

Fig. C3: Principle of balanced photodiodes: The original laser LO noise and signal S video noise are cancelled towards the IF output. The only noise resulting there is the power splitter noise at the shot noise limit. However, the beat between signal S and LO, the heterodyne intermediate frequency (IF) signal, is transmitted towards the IF output because of a phase shift of +90° and -90° at the power splitter towards PD1 and PD2, respectively.

A balanced photodiode subtracts the two optical noise signals after converting them to electrical signals:

$$\boldsymbol{\delta n_{bal}} := \boldsymbol{\delta n'} - \boldsymbol{\delta n''} = \sqrt{TR\,\bar{n}} \cdot 2\boldsymbol{\hat{n}_1} + (T-R) \cdot \boldsymbol{\delta n} \quad (B4)$$

The square of the power fluctuations shows that a balanced photodiode ($T = R = 0.5$) performs with a Fano factor of 1, if the efficiency of each photodiode is $\eta = 1$, which means that the laser power related noise output of the balanced photodiode is at the shot noise limit no matter how high the input Fano factor is.

$$|\delta n_{bal}|^2 := |\delta n' - \delta n''|^2 \quad (B5)$$
$$= 4(1-R)R\,\bar{n} + (1-2R)^2 \cdot \delta n^2$$
$$F_{bal} := \overline{|\delta n_{bal}|^2}/\bar{n} = 4(1-R)R + (1-2R)^2 \cdot F_{LO}$$

But the efficiency $\eta$ of photodiodes is in reality considerably less than 1, for example about $\eta = 0.75$ in our case. Then, a secondary noise propagation has to be considered, because the efficiency can be modeled as an absorber in front of an ideal photodiode:

$$\boldsymbol{\delta n_1''} = \sqrt{\eta(1-\eta)\,n_1'} \cdot \boldsymbol{\hat{n}_1'} + \eta \cdot \boldsymbol{\delta n_1'}$$

wherein

$$\boldsymbol{\delta n_1'} = -\sqrt{TR\,\bar{n}} \cdot \boldsymbol{\hat{n}_1} + T \cdot \boldsymbol{\delta n}$$
$$\bar{n}_1' = T\,\bar{n}$$

or together

$$\boldsymbol{\delta n_1''} = \sqrt{(1-\eta)\eta T\,\bar{n}} \cdot \boldsymbol{\hat{n}_1'} - \eta\sqrt{TR\,\bar{n}} \cdot \boldsymbol{\hat{n}_1} + \eta T \cdot \boldsymbol{\delta n} \quad (B6)$$

and on the other photodiode:

$$\boldsymbol{\delta n_2''} = \sqrt{(1-\eta)\eta R\,\bar{n}} \cdot \boldsymbol{\hat{n}_2'} + \eta\sqrt{TR\,\bar{n}} \cdot \boldsymbol{\hat{n}_1} + \eta R \cdot \boldsymbol{\delta n}$$

Now in total:

$$\overline{|\delta n_{bal}|^2} := \overline{|\delta n_1'' - \delta n_2''|^2} \quad (B7)$$
$$= (1-\eta)\eta(R+T)\bar{n} + 4\eta^2 TR\,\bar{n} + \eta^2(T-R)^2 \cdot \delta n^2$$

because $\boldsymbol{\hat{n}_1}$, $\boldsymbol{\hat{n}_1'}$ and $\boldsymbol{\hat{n}_2'}$ are all statistically independent, and so terms like $\overline{\boldsymbol{\hat{n}_1'} \cdot \boldsymbol{\hat{n}_2'}}$ are zero. Additionally, we have for a lossless power splitter ($R + T = 1$):

$$\overline{|\delta n_{bal}'|^2} = \quad (B8)$$
$$\eta[1 + \eta(4TR - 1)]\,\bar{n} + \eta^2(1-2R)^2 \cdot \delta n^2$$

or with $F_{bal} := \overline{|\delta n_{bal}'|^2}/\bar{n}$ (with the photon number taken before the beam splitter)

$$F_{bal} = \eta[1 + \eta(4TR - 1)] + \eta^2(1-2R)^2 \cdot F_{LO} \quad (B9)$$

The crucial but trivial assumption supporting the claim of this paper is now that the normalized random phasor $\boldsymbol{\hat{n}_A}$ generated at a power splitter $A$ at a balanced photodiode assembly $A$ is statistically independent from that generated at a second power splitter *B* in another balanced photodiode assembly *B*, $\hat{n}_B$. Then, the complex cross-correlation residual power between the post-detection noise voltages of the two receivers is, as in (39):

$$P_{CC}(\omega,t) := \frac{1}{Z}\overline{V_{AC,A}(\omega,t) \cdot V_{AC,B}^*(\omega,t)} \propto \quad (B10)$$

$$\overline{\delta n_{bal,A} \cdot \delta n_{bal,B}^*} = \eta^2(1-2R_A)(1-2R_B)F_{LO}\,\overline{n}$$

If phases are drifting or oscillating within the integration time or even the Fourier-transform time of the correlator, then we observe a loss of cross-correlation. This could not always be avoided in the current setup due to the inevitable long fibers. A solution would be to drastically reduce the optical path lengths by integration into a photonic chip.

The auto-correlation signal is not sensitive to a phase:

$$P_{AC,i}(\omega,t) := \frac{1}{Z}\overline{|V_{AC,i}(\omega,t)|^2} \propto \overline{|\delta n_{bal,i}|^2}$$
$$= \eta[1+\eta(4TR-1)] + \eta^2(1-2R)^2 \cdot F_{LO}\overline{n} \quad (B11)$$

**Appendix C**
**Optimum amplification to sample the laser noise**

In order to take into account the digitization noise of the ROACH, we project it towards the input of the amplifiers. To cancel out the laser noise, the digitization should sample the single receiver noise as precisely as possible. Therefore, one seeks a resolution (the number of bits) as high as possible. Additionally to that, there is the problem of optimizing the amplification $G$: a too high amplification runs the ADC into saturation, i.e. the largest noise peaks are cut off too often (and may actually damage the ADC). But if the largest peaks are not to be cut off to high probability, the majority of voltage swings is under-sampled. To find a trade-off between resolution and saturation, the outlier peaks (the wings of the Rayleigh probability distribution of the noise peaks) should not be cut off too often, say for example with just 1% probability. Then, according to B.M. Oliver [11] we have for a generalized amplitude A:

$$p(A)dA = \frac{A}{P_{av}}e^{-\frac{1}{2}\frac{A^2}{P_{av}}}dA \quad (C1)$$

and with $A = V/\sqrt{R}$ and $P_{av} = A^2/2 = A_{rms}^2$:

$$p(V)dV = \frac{V}{V_{rms}^2}e^{-\frac{1}{2}(\frac{V}{V_{rms}})^2}dV \quad (C2)$$

The cumulative distribution of this is

$$F(V) := \int_0^V p(V')dV' = 1 - e^{-\frac{1}{2}(\frac{V}{V_{rms}})^2} \quad (C3)$$

If we set thus $F(V) = 0.99$ we obtain $V_{max} = 2V_{rms}$, but if we set for example $V_{max} = 1.5V_{rms}$ we lose already 32% of the peaks.

To determine the optimal amplification for a certain laser power, we set the variance of the photocurrent fluctuations, and therefore the amplifier input voltage fluctuations over the complete ADC input bandwidth $\Delta\nu_S$, as proportional to the variance of the laser power fluctuations, which are in turn related to the laser power, $\delta P_{LO}^2 = Fh\nu P_{LO}\Delta\nu_S$. This relation stems from $\delta n_{LO}^2 = Fn_{LO}$ with $P_{LO} = h\nu\, n_{LO}/\Delta t$ and $\Delta\nu_S = 1/\Delta t$, if we take for example as $\Delta t$ the coherence time.

$$\sqrt{h\nu P_{LO}\Delta\nu_S} = \delta P_{LO} = \delta I_{rms,det}/\Re = (\delta V_{rms,det})/(Z\Re) = V_{rms,ADC}/(\sqrt{G}Z\Re) \quad (C4)$$

$G$ is the LNA-chain power amplification, $Z$ its input impedance, and $\Re$ the responsivity of the photodiode, and $\Delta t$ the averaging (integration) time. The double sideband bandwidth was for us $B = 2 \cdot 800$ MHz, since we are interested in the maximal voltage fluctuations due to the total bandwidth. The laser power is 1mW per photodiode. $\Re = \frac{\eta e}{h\nu} = \eta \cdot 1.257$ A/W, with $\eta = 0.75$. The iADC used has a maximum input voltage of 0.8V. The targeted rms-voltage is therefore $V_{rms} = 0.4$ V. The optimum gain is therefore

$$G = \frac{V_{rms}^2}{(Z\Re)^2 h\nu B P_{LO}} =$$

$$\frac{0.4^2}{(50 \cdot 0.75 \cdot 1.257)^2 \cdot 6.63 \cdot 10^{-34} \cdot 1.9 \cdot 10^{14} \cdot 1.6 \cdot 10^9 \cdot 10^{-3}}$$

$$= 85.5\, dB \quad (C5)$$

This is much more than we actually could realize (66.5 dB, 71.5 dB, and 76.5 dB) without running too much into saturation (which was clearly seen at 76.5 dB), so that the resolution of the laser shot noise seems to be improvable only by increasing the resolution of the ADC. A long time we did not test these limits for security reasons, but finally we made an investigation of the laser noise reduction over the amplification value (see Table 2). The ADC used has 8 bits, i.e. $2^8 = 256$ voltage digitization levels, so that a maximum laser noise to digitization noise power ratio of $\frac{\delta n_{LO}}{N_{dig}} \approx \left(\frac{V_{max}}{\Delta V_{dig}}\right)^2 = (2^{8-2})^2 = 4096 = 36$ dB may be achieved, where it is assumed that the digitization noise occurs over two bits.

**Appendix D: Proof of the Wiener-Khinchin theorem for cross correlation power**

This theorem proves that the Fourier transform

$$P_{CC}(\omega,t) := \int_{-\infty}^{\infty} d\tau\, e^{i\omega\tau} C(\tau,t) \quad (D1)$$

of the "original" or "time-lag" (XF-) correlation coefficient

$$C(\tau) := \int_{-\infty}^{\infty} dt'\, s_1(t') s_2(t' + \tau) \quad (D2)$$

is equivalent to FX-correlation, i.e. the simple multiplication of the Fourier-transforms of both signals. Because the integral over a random noise signal and its Fourier-transform do only exist over a finite time interval $T$, it is defined:

$$\tilde{s}_T(\omega) := \int_{-T/2}^{T/2} s(t) e^{-i\omega t} dt = \int_{-\infty}^{\infty} s(t) \prod_T(t) e^{-i\omega t} dt$$
$$=: \int_{-\infty}^{\infty} s_T(t) e^{-i\omega t} dt$$

using formally the box-function $\prod_T(t)$. With this it is:

$$\tilde{P}_{CC,T}(\omega) = \frac{\tilde{s}_{1,T}(\omega)\tilde{s}_{2,T}^*(\omega)}{T}$$
$$= \frac{1}{T} \int_{-\infty}^{\infty} \int_{-\infty}^{\infty} s_{1,T}(t_1)\, s_{2,T}^*(t_2) e^{-i\omega(t_1-t_2)} dt_1 dt_2$$

If we use therein the identity

$$s_{2,T}^*(t_2) = \int_{\infty}^{\infty} d\tau\, \delta(t_2 - t_1 - \tau) s_{2,T}^*(t_1 + \tau)$$

with

$$\delta(t_2 - t_1 - \tau) = \int_{-\infty}^{\infty} d\omega'\, e^{-i\omega'(t_2 - t_1 - \tau)}$$

giving

$$s_{2,T}^*(t_2) = \int_{\infty}^{\infty} d\tau \int_{-\infty}^{\infty} d\omega'\, e^{-i\omega'(t_2-t_1-\tau)}\, s_{2,T}^*(t_1+\tau)$$
$$= \int_{-\infty}^{\infty} d\omega' \int_{\infty}^{\infty} d\tau\, s_{2,T}^*(t_1+\tau) e^{-i\omega'(t_2-t_1-\tau)}$$

then we obtain:

$$P_{CC,T}(\omega) = \frac{1}{T} \int_{-\infty}^{\infty} \int_{-\infty}^{\infty} s_{1,T}(t_1)\, s_{2,T}^*(t_2) e^{-i\omega(t_1-t_2)} dt_1 dt_2$$
$$= \frac{1}{T} \int_{-\infty}^{\infty} \int_{-\infty}^{\infty} s_{1,T}(t_1) \int_{-\infty}^{\infty} d\omega' \int_{\infty}^{\infty} d\tau\, s_{2,T}^*(t_1 + \tau) e^{-i\omega'(t_2-t_1-\tau)}\, e^{-i\omega(t_1-t_2)} dt_1 dt_2$$
$$= \frac{1}{T} \int_{-\infty}^{\infty} d\omega' \int_{\infty}^{\infty} d\tau\, e^{+i\omega'\tau} \int_{-\infty}^{\infty} dt_1\, s_{1,T}(t_1) s_{2,T}(t_1 + \tau)\, e^{-it_1(\omega - \omega')} \int_{-\infty}^{\infty} dt_2\, e^{+it_2(\omega - \omega')}$$
$$= \frac{1}{T} \int_{-\infty}^{\infty} d\omega' \int_{\infty}^{\infty} d\tau\, e^{+i\omega'\tau} \int_{-\infty}^{\infty} dt_1\, s_{1,T}(t_1) s_{2,T}(t_1 + \tau)\, e^{-it_1(\omega - \omega')}\, \delta(\omega - \omega')$$
$$= \frac{1}{T} \int_{\infty}^{\infty} d\tau\, e^{+i\omega\tau} \int_{-\infty}^{\infty} dt_1\, s_{1,T}(t_1) s_{2,T}^*(t_1 + \tau)$$
$$=: \int_{\infty}^{\infty} d\tau\, C_T(\tau)\, e^{-i\omega\tau} =: \tilde{C}_T(\omega)$$

wherein

$$C_T(\tau) := \frac{1}{T} \int_{-\infty}^{\infty} dt_1\, s_{1,T}(t_1) s_{2,T}^*(t_1 + \tau)$$